\documentstyle[12pt]{article}
\def\baselinestretch{1.5}
\begin{document}
\pagestyle{empty}
\centerline{               \hfill NHCU--HEP--95--07}
\vskip 1cm
\centerline{\bf \large
On Exactly Solvable Potentials
}\vskip 1cm
\centerline{
Darwin Chang$^{(1,2)}$ and
We-Fu Chang$^{(1)}$ }\vskip 1cm
\centerline{\small\it
$^{(1)}$Physics Department,
National Tsing-Hua University, Hsinchu, Taiwan}
\centerline{\small\it
$^{(2)}$Institute of Physics, Academia Sinica, Taipei, Taiwan}
\date{\today}
\vfill
\begin{abstract}
We investigate two methods of obtaining exactly solvable
potentials with analytic forms.
\end{abstract}

\vskip 1in
\centerline{PACS numbers:03.65.Ge,11.30.Pb  \hfill}

\newpage

\pagestyle{plain}

 There are two methods of obtaining exactly solvable potentials in quantum
mechanics.  The first method was developed by applying the technique of
supersymmetry (SUSY) to the Schr\"odinger equation and obtain two
potentials with almost identical spectra.  The two potentials can be
considered to be
superpartners of each other.  It has been shown \cite{genden,table} that
if the two partners happen to be related by a simple relationship called shape
invariance, the energy eigenvalues of the potential can be solved exactly.
All the known solvable potentials with closed analytic forms can be shown
to be shape invariant.  In the literature, there are other
solvable potentials that have not been shown to be shape invariant,
however, they exist only complex numerical forms which we shall not
consider in this article.

A second interesting method of obtaining solvable potential was
proposed by Klein and Li\cite{klein} based on some special
quantum commutation relationships.  Li\cite{li} has recently
worked out the most general potential that can be obtained this way.

In this paper, we first investigate the relationship between the
two approaches.  We will show that the general solutions obtained by Li are
special cases of the general solutions that can be obtained
by solving shape invariance condition.  Therefore, the solving shape
invariance condition remains the most general method of obtaining exactly
solvable potentials which can be given in analytic form.
Unfortunately, there is no general method for
getting analytic solutions of shape invariance condition.  The best one
can do seems to be starting from a guessing ansatz with an unknown
function.  The shape invariance is then enforced by demanding that the
function satisfies an ordinary differential equation.  The ansatz allows one
to turn the difficult shape invariance condition into a problem of
solving differential equation.
It also allows one to associate each ansatz as defining a particular
class of solutions.  While
the solutions obtained by Li correspond to those defined by a particular
ansatz, in the literature, the largest classes of analytic solutions of
the shape invariance condition was provided by Gendenshtein
\cite{genden}.  He proposed three ansatzs which define three classes of
solutions which seem to cover all the known analytic exactly
solvable potentials.
We, therefore, proceed to solve the differential equations corresponding
to these ansatzs and
the energy eigenvalues of these three classes of solutions of shape
invariance condition.  In the process, we also
demonstrate that all the known solutions in the literature are special cases
of the one provided by Gendenshtein.

The three ansatzs of Gendenshtein and the forth one corresponding to Li's
solutions also demonstrated that each potential can be
simultaneously represented in many classes.  It is also clear that one
can in principle continue to invent more ansatz without obtaining any new
potentials.  Therefore, one is immediately
faced with the intriguing question of what is the better way to classify
these solvable potentials than using the ansatzs.  One method seems to be
starting from the $n$ dependence of its spectrum.  It is noted that only
a few simple
general forms of $n$ dependence are allowed for the known exactly solvable
potentials.  We shall make some comments in this direction later and mention
some interesting unsolved problems in the conclusion.

We  first briefly review
the procedure of constructing creation and annihilation operators
of simple harmonic oscillator, now we can apply this method to the
Schr\"odinger
equation with an arbitrary potential $V(x)$.
\begin{equation}
(-\frac12 \frac{d^2}{dx^2}+V(x))\psi=E\psi \nonumber
\end{equation}
Assuming that we have already shifted the potential by a constant so that
the ground state energy becomes zero, we shall denote this Hamiltonian and
its potential with $(-)$ subscript and denote $\psi_n^{(-)}$
as its eigenfunctions.  That is,
$V_{-}(x)\equiv  V(x)$,
$\psi_0^{(-)}(x)\equiv \psi_{0}(x)$,
$E_0^{(-)} = 0$,
with
\begin{equation}
H_{-}\psi_0^{(-)}=(-\frac{1}{2} \frac{d^2}{dx^2}+V_{-}(x))\psi_0^{(-)}=0.
\label{eqn:w531} \end{equation}
The above equation, Eq.(\ref{eqn:w531}) is identical to
\begin{equation}
H_{-}=\frac{1}{2}(-\frac{d^2}{dx^2}+\frac{\psi^{(-)''}_0}{\psi_0^{(-)}}).
\end{equation}
Then, the general creation and annihilation operators are:
\begin{eqnarray}
A^{+}&=&\frac{1}{\sqrt{2}}
(-\frac{d}{dx}-\frac{\psi^{(-)'}_0}{\psi^{(-)}_0}),\nonumber\\
A &=&\frac{1}{\sqrt{2}} (\frac{d}{dx}-\frac{\psi^{(-)'}_0}{\psi^{(-)}_0}).
\end{eqnarray}
The Hamiltonian can be written as $H_{-}=A^{+}A$.
Now we define a new Hamiltonian $H_{+}=AA^{+}$.
which can be written as
$H_{+} \equiv -\frac{1}{2}\frac{d^2}{dx^2}+V_{+}(x).$
The corresponding potential in the new Hamiltonian is:
\begin{equation}
V_{+}= - V_{-}+(\frac{\psi_{0}^{(-)'}}{\psi^{(-)}_0})^2
= V_{-}-\frac{d^2}{dx^2} \ln \psi_0^{(-)}.
\label{eqn:v+v-}
\end{equation}
The $V_{+}$, $V_{-}$ are called supersymmetric partner potentials.

One can define a new function, $W(x)$, called superpotential,
\begin{equation}
W(x)=-\frac{1}{\sqrt{2}}\frac{\psi_0^{(-)'}}{\psi^{(-)}_0}.
\end{equation}
Solving the differential equation we get
\begin{equation}
\psi^{(-)}_0(x)=exp(-\sqrt{2}\int^{x} W(x') dx'),
\end{equation}
and $A^{+} = -\frac{1}{\sqrt{2}}\frac{d}{dx}+W(x)$, $A =
\frac{1}{\sqrt{2}}\frac{d}{dx}+W(x)$.
The two partner potentials are made from superpotential $W(x)$,
\begin{equation}
V_{\pm}=W^2 \pm \frac{1}{\sqrt{2}} \frac{d}{dx} W(x).
\end{equation}
$A^+$ and $A$ do not commute with each other and satisfy
$\lbrack A,A^{+} \rbrack =\sqrt{2} W(x)^{'}$.
To compare the spectra of these two partners,
denote $\psi^{(-)}$ as the eigenfunction of $H_{-}$ and $\psi^{(+)}$ as
that of $H_{+}$.
Then, $A\psi^{(-)}$ is an eigenfunction of $H_{+}$ because
\begin{equation}
H_{+}(A\psi^{(-)}_{n})=AA^{+}A\psi^{(-)}_{n}=AH_{-}\psi^{(-)}_{n}
=E^{(-)}_{n}(A\psi^{(-)}_{n}).
\end{equation}
Similarly, $A^+ \psi^{(+)}$ is an eigenfunction of $H_{-}$,
$H_{-}(A^{+}\psi^{(+)}_{n}) = E^{(+)}_{n}(A^{+}\psi^{(-)}_{n})$.
Therefore $A$, $A^+$ can transform an eigenfunction of one potential into a
eigenfunction of it's partner with the same energy.
However, note that the ground state of $V_{-}$ is annihilated by $A$ and
have no partner state of $V_{+}$.  Therefore the ground state of $V_{-}$
has no superpartner.

Relations of energy eigenvalues and wavefunctions are:
\begin{eqnarray}
E^{(+)}_{n} &=& E^{(-)}_{n+1},\nonumber \\
\psi^{(+)}_{n} &=& \frac{1}{\sqrt{E^{(-)}_{n+1}}}A\psi^{(-)}_{n+1} \mbox{,\ \ \
}n=0,1,2,\ldots,\nonumber\\
\psi^{(-)}_{n+1} &=& \frac{1}{\sqrt{E^{(+)}_{n}}}A^{+}\psi^{(+)}_{n} \mbox{,\ \
\
}n=0,1,2,\ldots.
\label{eqn:poi087}
\end{eqnarray}

In 1983, Gendenshtein \cite{genden} suggested that all exactly
solvable
potentials are ``shape invariant" ( he used ``form invariant" instead ).
Shape invariance means that the superpartner potentials have the same
$x-$dependence modulo the changes in values of a set of
parameters that define the potential.
Mathematically, it means the superpartner of $V_{-}(x;a_{0})$,
$V_{+}(x;a_{0})$ can be written as
\begin{eqnarray}
V_{+}(x;a_{0})
= V_{-}(x;a_{1})+R(a_1)
\label{eqn:huhq897}
\end{eqnarray}
and $a_1 = f(a_0)$,
where $a_0$ are the set of parameters in $V_{+}$, and $f$ is the
transformation function that
maps $a_0$ into $a_1$.  The remainder $R(a_1)$ is
independent of $x$.

To show that the eigenvalues can be obtained easily from the above
condition, we
construct a series of Hamiltonian $H^{(k)}$, $ k=0,1,2,\ldots $, with
$H^{(0)} \equiv H_{-}$,
$H^{(1)} \equiv H_{+}$,
\begin{equation}
H^{(k)} = -\frac{1}{2}\frac{d^2}{dx^2}+V_{-}(x;a_{k})+\sum^{k}_{s=1}
R(a_{s}),
\label{eqn:hk}
\end{equation}
where $a_{s}=f^{s}(a_0) $, i.e., $f$ mapped $s$ times. Furthermore, we have
{}From Eq.(\ref{eqn:hk}), we see that
the ground state of $H^{(k)}$ has the energy:
\begin{equation}
E_0^{(k)}=\sum^{k}_{s=1}R(a_{s}).
\end{equation}
Since $(n+1)$th energy eigenvalue of
$H^{(0)}$ (=$H_{-}$), whose ground state energy is zero,
is coincident with the ground state energy of
Hamiltonian $H^{(n)}$, the complete eigenvalues of $H_{-}$ are:
\begin{equation}
 E^{(-)}_{n}=\sum^{n}_{k=1}R(a_{k}) \mbox{\ \ ,\ \ } E^{(-)}_0=0.
\end{equation}

If a potential is shape invariant, we can also get the bound state
wavefunctions $\psi^{(-)}_{n}$ easily. This is because  $A$ and $A^{+}$ can
link up the wavefunctions of the superpartners with the same energy.
Starting from $H^{(-)}_{n}$, it's ground state $\psi_0^{(-)}(x;a_{n})$
corresponds to the first excited state $\psi_1^{(-)}(x;a_{n-1})$ of
$H^{(-)}_{n-1}$.  In the same manner, eventually,
it will correspond to the $n$th state of $H^{(-)}_0$.
Recalling Eq.(\ref{eqn:poi087}), one obtains
\begin{eqnarray}
\psi^{(-)}_{n} \propto A^{+}(x;a_0)A^{+}(x;a_1)\ldots
 A^{+}(x;a_{n-1})\psi^{(-)}_0(x;a_{n}).
\end{eqnarray}

To compare the supersymmetry approach with the second approach
proposed by Klein and Li\cite{klein,li}, we shall first give a
brief review.
Consider an one dimensional quantum system with the Hamiltonian
\begin {equation}
H=-\frac12\frac{d^2}{dx^2}+V(x).
\end{equation}
For an arbitrary function of position, $f=f(x)$, one can easily derive
\begin{equation}
\lbrack\lbrack f,H\rbrack,H\rbrack = -(f^{\prime\prime}H+Hf^{\prime\prime})
+2f^{\prime\prime}V+f^{\prime}V^{\prime}-\frac{f^{(4)}}4.
\end{equation}
To obtain solvable potential Klein and Li proposed to impose
a so-called linear double commutator relation demanding that for
some $f$ and $V$, the right hand side become a linear functional of $f$.
In order to make the operator equation linear in $f$, one imposes
\begin{equation}
2f^{\prime\prime}V+f^{\prime}V^{\prime}=\alpha f+\beta,
\label{eqn:jh23478}
\end{equation}
and
\begin{equation}
f^{\prime \prime}=\mu f+\nu.
\label{eqn:t565893}
\end{equation}
Eq.(\ref{eqn:jh23478}) relates $V$ to $f$ and can be solved
 to give the potential $V(x)$
\begin{equation}
V(x)=\frac{\alpha f^2+2\beta f+\gamma}{2(f^{\prime})^2}.
\label{eqn:vvv}
\end{equation}
Eq.(\ref{eqn:t565893}) can be solved for $f$ as
\begin{equation}
f=a x^2+b x+c\mbox{\ \ \ \ \ ( $\mu=0$ ),}
\end{equation}
or
\begin{equation}
f=A e^{\sqrt{\mu}x}+B e^{- \sqrt{ \mu}x}+C
\mbox{\ \ \ \ \ ( $\mu\not= 0$ ).}
\end{equation}
For $\mu = 0$, the resulting potentials are just those for one- or
three-dimensional harmonic oscillator problems.
For $\mu\not= 0$, one can get the general potential
\begin{equation}
V(x)=\frac{\beta(Ae^{\sqrt{\mu}x}+B e^{-\sqrt{\mu} x})+\frac{\gamma}{2}}
{\mu(A e^{\sqrt{\mu} x}-B e^{-\sqrt{\mu} x})^2},
\end{equation}
which are
special cases of Morse and P\H oschl-Teller potentials.
Li also solved the energy eigenvalues by the method of Heisenberg's matrix
mechanics\cite{li} under some general assumptions.
The eigenvalue is proportional to $n^2$, where $n$ is a
quantum number.  Also, from
Eqs. (\ref{eqn:t565893}) and (\ref{eqn:vvv}),
the corresponding superpotential, with
linear double commutator relation, can be written as\cite{li}
\begin{equation}
W(x)=\frac{a f(x)+b}{f^{\prime}(x)}.
\label{eqn:LiSI}
\end{equation}
Here we shall first show that the superpotential is indeed shape
invariance and then solve the eigenvalues using the supersymmetric
methods outlined earlier.

Given, Eqs.(\ref{eqn:LiSI}) and (\ref{eqn:t565893}), one can easily check the
shape invariance
condition by working out the potentials $V_{\pm}$ as
\begin{equation}
V_{\pm}=\frac{a(a\mp \frac{\mu}{\sqrt{2}})f^2+(a(b\mp \frac{\nu}{\sqrt{2}}
)+b(a\mp \frac{\mu}{\sqrt{2}}))f+b(b\mp\frac{\nu}{\sqrt{2}})}{ (f^{\prime})^2}
\pm \frac{a}{\sqrt{2}}.
\end{equation}
Therefore, $V_{\pm}$ are related by shape invariance condition
\begin{equation}
V_{+}(a,b;x)=V_{-}(a-\frac{\mu}{\sqrt{2}},b-\frac{\nu}{\sqrt{2}}
;x)+\sqrt{2}a-\frac{\mu}{2}.
\end{equation}
The energy eigenvalues can be
straight-forwardly worked out to be
\begin{equation}
E_{n}=\sqrt{2}n a-\frac{\mu n^2}{2},
\end{equation}
where the ground state energy has been shifted to zero.  SUSY method
and Li's method gives the same results as expected.

In the SUSY approach,
shape invariance requires that the superpotential $W(x)$
satisfies the functional differential equation:
\begin{equation}
W^2(a,x) + \frac{1}{\sqrt{2}}W^{\prime}(a,x) = W^2(a_{1},x) -
\frac{1}{\sqrt{2}}W^{\prime} (a_1,x) +R (a_1), \label{eqn:SI}
\end{equation}
where $a_0$ represents a set of parameters,
called the shape invariance relation(SIR).
The values of the set, $a_1$,
on the right-hand side, depend on the value of $a_0$, that is, $a_1=f(a_0)$
for some function $f$.
The only known way of solving SIR is to impose an educated ansatz to turn
it into a differential equation.  In this direction,
Gendenshtein \cite{genden} had proposed three ansatzs which provide some
general classes of solutions.
There is no claim that the three classes should
encompass the most general solutions of SIR.
However, it is interesting to note that all the known solutions can be shown
to be special cases of one of the three classes as we shall demonstrate
later.   Also, the three classes are not mutually exclusive.
That is, some solutions can be represented in more than one classes.

We shall discuss the three classes in order, solve the corresponding
nonlinear differential equations and work
out the potentials and energy eigenvalues for each class.
The three classes can be described as follow:\\
(I).Class one:\\
The ansatz for the superpotential is of the form
\begin{eqnarray}
W=a f_1+b.
\end{eqnarray}
The SIR then requires $f_1$ to satisfy
\begin{eqnarray}
f^{\prime}_1=p f_1^2+q f_1+r.
\end{eqnarray}
$V_{\pm}$ can be worked out to be
\begin{equation}
V_{\pm}=a(a\pm \frac{p}{\sqrt{2}})f^2_1+2a(b \pm \frac{q}{2\sqrt{2}})f_1
       + b^2\pm \frac{a r}{\sqrt{2}},
\end{equation}
with the parameters transforming as
\begin{eqnarray}
&& a_n=a+\frac{n p}{\sqrt{2}}, \nonumber \\
&& b_n=\frac{a b}{a+\frac{n p}{\sqrt{2}}}+\frac{n a \frac{q}{\sqrt{2}}
        + \frac{n^2 p q}{4}}{(a+\frac{n p}{\sqrt{2}})},\\ \nonumber
&& R=b_0^2-b_1^2+\frac{r}{\sqrt{2}}(a_0+a_1).
\end{eqnarray}
For the case when $p\not= 0$, the eigenvalues can be worked out to be
\begin{eqnarray}
E_n&=&\sum_{k=1}^{n} R(a_k,b_k) \nonumber \\
&=&\sum_{k=1}^{n}\lbrack
(b_{k-1}^2-b_{k}^2)+\frac{r}{\sqrt{2}}(a_{k-1}+a_{k})\rbrack\nonumber \\
&=&b^2-\lbrack \frac{a b}{a+\frac{n p}{\sqrt{2}}}+\frac{n a
\frac{q}{\sqrt{2}}+\frac{n^2 p q}{4}}{(a+\frac{n p}{\sqrt{2}})}
\rbrack^2+\sqrt{2} a r n +\frac{p r}{2} n^2.
\end{eqnarray}
Ordering the terms by power of $n$, $E_{n}$ can be written as
\begin{equation}
E_{n}=(b-\frac{a q}{2p})^2
-(b-\frac{aq}{2p})^2\frac{a^2}{(a+n\frac{p}{\sqrt{2}})^2}+
\sqrt{2}(a r-\frac{a q^2}{4p})n+(\frac{pr}{2}-\frac{q^2}{8})n^2.
\end{equation}
For the case when $p = 0$, the transformation of parameter and the
eigenvalues can be obtained as
\begin{eqnarray}
b_{n}&=&b+\frac{n q}{\sqrt{2}}, \\ \nonumber
R&=&b_0^2-b_1^2+\sqrt{2}a r,\\ \nonumber
E_n&=&\sum^{n}_{k=1}R_{k}=\sqrt{2}(a r-b q)n-\frac{q^2}{2}n^2.
\end{eqnarray}
The known exactly solvable potentials in this class
and their corresponding parameters are listed in Table 1.

\begin{table}[hbt]
\begin{center}
\caption{\ \ Class One \ \ \ $W=a f(x)+b$\ ;\ $f^{'}=pf^2+qf+r$}
\begin{tabular}{|l|lcccccc|}\hline
Potential& $W(x)$ & $f(x)$ & $a$ & $b$ & $p$ & $q$ & $r$\\
\hline
&&&&&&&\\
Shifted Oscillator&$\sqrt{\frac12}\omega x-b
$&$x$&$\sqrt{\frac12}\omega$&$-b$&$0$&$0$&$1$\\
&&&&&&&\\
Coulomb&$\sqrt{\frac12}\frac{e^2}{l+1}-\frac{l+1}{\sqrt{2}
r}$&$r^{-1}$&$-\frac{l+1}{\sqrt{2}}$&$\sqrt{\frac12}\frac{e^2}{l+1}$&$-1$&$0$&$0$\\
&&&&&&&\\
Morse&$A-Be^{-\alpha x}$&$e^{-\alpha x}$ & $-B$ & $A$ & $0$ &$-\alpha$&$0$\\
&&&&&&&\\
Rosen-Morse&$A\tanh \alpha x +\frac{B}{A}$&$\tanh\alpha
x$&$A$&$\frac{B}{A}$&$-\alpha$&$0$&$\alpha$\\
&&&&&&&\\
Eckart&$-A\coth\alpha r+\frac{B}{A}$&$\coth \alpha
r$&$-A$&$\frac{B}{A}$&$-\alpha$&$0$&$\alpha$\\
&$(B>A^2)$&&&&&&\\ \hline
\end{tabular}
\end{center}
\end{table}

The most general solutions for function $f$ in this case can be
summarized as
\begin{eqnarray}
&& f=r x+c \mbox{\ \ }, (p=q=0 );\\
&& f=-\frac{1}{p x+c}\mbox{\ \ }, (p\not=0,q=r=0 );\\
&& f=c e^{q x}-\frac{r}{q}\mbox{\ \ }, (p=0, q\not=0 );\\
&& f=\frac{\sqrt{4p r-q^2}}{2p}
\tan(\frac{\sqrt{4p r-q^2}}{2}x+c)-\frac{q}{2p}\mbox{\ \ },
(p\not=0, q\mbox{\ or\ } r\not=0 ),
\end{eqnarray}
where $c$ is an integration constant.

\noindent
(II). Class two:\\
The superpotential in this ansatz is assumed to be of the form
\begin{eqnarray} &&W=a f_2+\frac{b}{f_2}, \nonumber \\
&&f^{\prime}_2=p f^2_2+q.
\end{eqnarray}
$V_{\pm}$ can be worked out to be
\begin{equation}
V_{\pm}=a(a\pm \frac{p}{\sqrt{2}})f^2_2+b(b\mp\frac{q}{\sqrt{2}})f^{-2}_2
        + 2a b\pm\frac{1}{\sqrt{2}}(a q-p b),
\end{equation}
with the parameters transforming as
\begin{eqnarray}
&&a_n=a+\frac{n p}{\sqrt{2}}, \nonumber \\
&&b_n=b-\frac{n q}{\sqrt2}, \nonumber\\
&&R_n=2\sqrt{2}(a q-b p)+2(2n-1)p q.
\end{eqnarray}
The eigenvalues are
\begin{equation}
E_n=2\sqrt{2}(a q-b p)n+2p q n^2.
\end{equation}
The well-known examples in the class and their corresponding parameters
are listed in Table 2 for illustration. 

\begin{table}[hbt]
\begin{center}
\caption{\ \ Class Two\ \ \ $W=a f+\frac{b}{f}$\ ;\ $f^{'}=pf^2+q$}
\begin{tabular}{|l|lccccc|}\hline
Potential& $W(x)$ & $f(x)$ & $a$ & $b$ & $p$ & $q$ \\
\hline
&&&&&&\\
3-D Oscillator&$\sqrt{\frac12}\omega r-\frac{l+1}{\sqrt{2}r}
$&$r$&$\sqrt{\frac12}\omega$&$-\frac{l+1}{\sqrt{2}}$&$0$&$1$\\&&&&&&\\

P\"oschl-Teller I& $A\tan\alpha x-B\cot\alpha x$ & $\tan\alpha x$
&$A$&$-B$&$\alpha$&$\alpha$\\
&$(0<\alpha x<\frac{\pi}{2})$&&&&&\\
P\"oschl-Teller II& $A\tanh\alpha r-B\coth\alpha r$ & $\tanh\alpha r$
&$A$&$-B$&$-\alpha$&$\alpha$\\
&$(B<A)$&&&&&\\ \hline
\end{tabular} \end{center}
\end{table}

The general solutions for function $f$ in this case can be written as
\begin{eqnarray}
&& f=qx+c\mbox{\ \ }( p=0 ),\\
&& f=\sqrt{\frac{q}{p}}\tan (\sqrt{pq} x+c)\mbox{\ \ }( p\not=0,q\not=0 ),\\
&& f=-\frac{1}{p x+c}\mbox{\ \ }( p\not=0,q=0 ),
\end{eqnarray}
where $c$ is an integration constant.

\noindent
(III).Class three:\\
The superpotential is assumed to be of the form
\begin{eqnarray}
&&W=\frac{a+b\sqrt{p f^2_3+q}}{f_3},\nonumber \\
&&f^{\prime}_3=\sqrt{p f^2_3+q}.
\end{eqnarray}
$V_{\pm}$ can be worked out to be
\begin{equation}
V_{\pm}=\frac{1}{f_3^2} \lbrack a^2+b q(b\mp\sqrt{\frac12})
        + 2a(b\mp \frac{1}{2\sqrt{2}})\sqrt{p f^2_3+q} \rbrack +b^2 p,
\end{equation}
with the parameters transforming as
\begin{eqnarray}
&&b_n=b-\frac{n}{\sqrt2}, \nonumber\\
&&R_n=p(\sqrt{2}b+\frac12)-pn.
\end{eqnarray}
The eigenvalues are
\begin{equation}
E_n=\sqrt{2}n p b-\frac{p n^2}{2}.
\end{equation}
The well-known exactly solvable potentials in this class are listed in
Table 3.

The general solutions for function $f$ are
\begin{eqnarray}
& & f=\sqrt{q}x+c \mbox{\ \ }, ( p=0 );\\
& & f=c e^{\sqrt{p} x}\mbox{\ \ }, ( p\not=0, q=0 );\\
& & f=\frac12\sqrt{\frac{q}{p}}
 (e^{\sqrt{p}(x+c)}-e^{-\sqrt{p}(x+c)})\mbox{\ \ }, ( p,q\not=0 ),
\end{eqnarray}
c is an integration constant.

\begin{table}[hbt]
\begin{center}
\caption{\ \  Class Three \ $W=(a+b\sqrt{pf^2+q})/f$ ,$f^{'}=\sqrt{pf^2+q}$ }
\begin{tabular}{|l|lccccc|}\hline
Potential& $W(x)$ & $f(x)$ & $a$ & $b$ & $p$ & $q$ \\
\hline
&&&&&&\\
Morse&$A-Be^{-\alpha x}$&$e^{\alpha x}$ &
$-B$&$\frac{A}{\alpha}$&$\alpha^2$&$0$\\ &&&&&&\\
&$A\tanh\alpha x+B\sec \!{\rm h} \alpha x$&$\cosh\alpha x$ &
$B$&$\frac{A}{\alpha}$&$\alpha^2$&$-\alpha^2$\\ &&&&&&\\

Rosen-Morse&$A\coth \alpha r-B\csc\!{\rm h}\alpha r$&$\sinh\alpha
r$&$-B$&$\frac{A}{\alpha}$&$\alpha^2$&$\alpha^2$\\
&$(A<B)$&&&&&\\
&&&&&&\\

Eckart&$-A\cot\alpha x+B\csc\alpha x$&$\sin\alpha
x$&$B$&$-\frac{A}{\alpha}$&$-\alpha^2$&$\alpha^2$\\
&$(0<\alpha x<\pi, A>B)$&&&&&\\ \hline
\end{tabular}
\end{center}
\end{table}
The tables showed that all known exact solvable potentials with analytic
forms can be put into one of these three classes.
Whatever has spectrum 1/$n^2$ can be classified as the first
case.
However, only the class two solutions can produce
three dimensional oscillator, type-I P\H oshl-Teller or
type-II P\H oshl-Teller potentials.
But two potentials, Morse and Eckart, can be considered both as case
one and as case three.
As mentioned before, Li's results are just the special case of shape
invariance solutions.
However it cannot be so easily fit into
one of the three classes given by Gendenshtein.
In fact one can consider
Eq.(\ref{eqn:t565893}) and Eq.(\ref{eqn:LiSI})
to be the equations that define a fourth ansatz for solutions of SIR.
The set of solutions overlaps with those of the other three classes
provided by Gendenshtein but does not generate a new one.

{}From this point of view, it is clear that the ansatz does not provide a very
precise classification of the solutions of shape invariance condition.
It is not too hard to propose new ansatz, however, it is much harder to
generate new solutions.
Typically one can not be sure whether an ansatz generates any new
solution or not until they are solved completely.
Therefore it seems that a better classifying solutions may be to use
the energy spectrum and its quantum number dependence instead.

   In conclusion, we have discussed the two methods of obtaining exactly
solvable potentials in quantum mechanics.
One of them requires the shape invariance between the
   superpartners of the potentials in the supersymmetric formulation.
The other one imposes a
so-called ``linear double commutator relations".
   We have shown that the second case only produces solutions which are
special solutions of the first approach.
{}From this point of view, the shape invariance approach is
still the most general method of producing the analytic, exactly solvable
potentials.  We also argued that the $n$ dependence of the energy spectrum
may be a better way of
   telling the difference between different classes of potentials.
   In addition, we work out the energy eigenvalues of
the most general classes of potentials in the literature.

Unfortunately, it is still not possible to obtain the most general
solutions to SIR.
{}From the table of \cite{table}, one observes that all the
known shape invariant potentials have basically only one parameter changing
under the SUSY transformation.  In particular, for P\"oschl-Teller I
potentials, only the combination $A+B$ is changing.
The parameter $A-B$ is invariant.
For P\"oschl-Teller potentials II, on the other hand,
the combination $A-B$ is changing and the parameter $A+B$ is invariant.
It is not surprising because as long as the transformation property of
the parameters are linear for a proper choice of parameters, one can
always make linear combinations such that only one of them is changing
during the transformation.
Since, in general, there is no reason for the transformation of the
parameters to be linear, one would expect to have a lot more interesting
solutions of shape invariance condition waiting to be discovered.

Another interesting observation is that all the shape invariant solutions
has the spectrum of one of the following the forms (modulo a constant that
sets the ground state energy to zero): (1) $a n$, ($a>0$); (2) $-{b\over
(n + a)^2}$, ($b>0$); (3) $\pm(a + b n)^2$ or their linear combination.
The harmonic oscillator is an example for the first form.  The hydrogen
atom is an example for the second form and, the square well is the
simplest example for the third form.  Also note that the general spectra
of all three classes of solutions suggested by Gendenshtein are all linear
combinations of the above three forms.  One may wonder if there is
something fundamental about these kinds of spectra that made them
represent the spectra of all the shape invariant, exactly solvable,
potentials.

Finally, regarding the approach proposed by Klein and Li, since Li's
solutions in \cite{li} produce only part of the solutions of SIR, it
suggests that it may be possible to generalize their approach within its'
framework.  We have made some attempts in this direction, however, so far
without success.

This work was supported in part by National Science Council of
Republic of China grants
No. NSC 84-2112-M-007-042 and
No. NSC 84-2112-M-007-016.

\renewcommand\baselinestretch{1}
\end{document}